\documentclass[aip,jap,reprint]{revtex4-1}
\usepackage{graphicx}
\DeclareGraphicsExtensions{.pdf,.jpg,.png,.eps}
\usepackage{amsfonts}
\usepackage{amsmath}
\usepackage{amssymb}
\usepackage{color}
\usepackage{color}%\usepackage{lineno}
\usepackage{epstopdf}

\begin{document}
%\preprint{}
\title{Spin-Reorientation and Weak Ferromagnetism in Antiferromagnetic TbMn$_{0.5 }$Fe$_{0.5 }$O$_3$}
\author{Hariharan Nhalil.$^*$}
\email{hariharan@physics.iisc.ernet.in, hariharan.nhalil@gmail.com}
\affiliation{Department of Physics, Indian Institute of Science, Bangalore 560012, India}
%\thanks{}
\author{Harikrishnan S. Nair}
\affiliation{Highly Correlated Matter Research Group, Physics Department, University of Johannesburg, P. O. Box 524, Auckland Park 2006, South Africa}
\author{Sanathkumar R.}
\affiliation{Department of Physics, Indian Institute of Science, Bangalore 560012, India}
\author{Andr\'{e} M. Strydom}
\affiliation{Highly Correlated Matter Research Group, Physics Department, University of Johannesburg, P. O. Box 524, Auckland Park 2006, South Africa}
\affiliation{Max Planck Institute for Chemical Physics of Solids (MPICPfS), N\"{o}thnitzerstra{\ss}e 40, 01187 Dresden, Germany}
\author{Suja Elizabeth}
\affiliation{Department of Physics, Indian Institute of Science, Bangalore 560012, India}
%\noaffiliation is required (may also be used with the \author command).
\date{\today}
\begin{abstract}
Orthorhombic single crystals of TbMn$_{0.5 }$Fe$_{0.5 }$O$_3$ are found to exhibit spin-reorientation, magnetization reversal and weak ferromagnetism. Strong anisotropy effects are evident in the temperature dependent magnetization measurements along the three crystallographic axes $\bf a$, $\bf b$ and $\bf c$. A broad magnetic transition is visible at $T^\mathrm{Fe/Mn}_N $ = 286 K due to paramagnetic to $A_xG_yC_z$  ordering. A sharp transition is observed at $T^\mathrm{Fe/Mn}_{SR} $ = 28~K, which is pronounced along {\bf c} axis in the form of a sharp jump in magnetization where the spins reorient to $G_xA_yF_z$ configuration. The negative magnetization observed below $T^\mathrm{Fe/Mn}_{SR}$ along  $\bf c$ axis is explained in terms of domain wall pinning. A component of weak ferromagnetism is observed in field-scans along $\bf c$-axis but below 28 K. Field-induced steps-like transitions are observed in hysteresis measurement along $\bf b$ axis below 28~K. It is noted that no sign of Tb-order is discernible down to 2~K. TbMn$_{0.5 }$Fe$_{0.5 }$O$_3$ could be highlighted as a potential candidate to evaluate its magneto-dielectric effects across the magnetic transitions.
\end{abstract}
%Use showkeys class option if keyword
\maketitle
\section{introduction}
\indent 
 $R$FeO$_3$ and $R$MnO$_3$ ($R$ = rare earth) perovskites exhibit multiferrocity\cite{tokunaga2008magnetic,kimura2003magnetic}, spin-switching\cite{jeong2012temperature,yuan2013spin}, magneto-optic\cite{de2011laser} and magneto-electric\cite{mandal_prl_107_2011spin} properties which are desirable in device application. It is only natural to try and combine the most desirable features of these two classes of compounds in composite solid solution.  The magnetic properties of multiferroic $R$MnO$_3$ ($R$ = Tb, Dy and Gd) are well documented\cite{kimura2003magnetic, kimura_prb_71_2005magnetoelectric}. It is now well established that switchable ferroelectric polarization in $R$MnO$_3$ results from the incommensurate magnetic order. On the other hand, in orthorhombic $R$FeO$_3$, non-collinear antiferromagnetic (AFM) $G$-type order of Fe$^{3+}$ leading to weak-ferromagnetism (WFM) is observed at high temperature ($\gtrapprox$ 300~K) while the $R$ sub-lattice orders antiferromagnetically at very low temperature ($\lessapprox$ 10~K)\cite{bellaiche2012simple,PhysRevLett.1.3}. Orthoferrites are also known to display spin-reorientation transitions. For instance DyFeO$_3$  shows Morin transition\cite{morin_PhysRev.78.819.2} where Fe$^{3+}$ spin system reorients from AFM configuration along $\bf a$ axis with WFM component along $\bf c$ axis to simple AFM along $\bf b$ axis at$\sim$35~K\cite{prelorendjo_jpc_13_1980spin}. The spin-transitions are advantageous when considering the magneto-electric properties because the anisotropic magnetic properties and lattice strain can be coupled in a ferroelectric material. 
\\
\indent
In TbFeO$_3$, 4$f$-electron based Tb sub-lattice and 3$d$-electron based Fe sub-lattices are coupled in anti-parallel manner.  Fe$^{3+}$ moments possess $\Gamma_4$ ($G_xA_yF_z$) order with $T^\mathrm{Fe}_N \approx$ 650~K\cite{bouree1975mise,bertaut1967structures,tejada1995quantum} accompanied by spin re-orientation to $\Gamma_2$ ($G_zC_yF_x$).  At 3 K, another spin re-orientation occurs and the Fe$^{3+}$ moments adopt $\Gamma_4$ ($G_xA_yF_z$) structure again. In the 10 - 3 K interval, the Tb$^{3+}$ moments are ordered in $\Gamma_2$ ($F_xC_y$) spin structure. Below 3 K, Tb$^{3+}$ order cooperatively in $\Gamma_8$ ($A_xG_y$) configuration. TbMnO$_3$, on the other hand is a well known multiferroic \cite{kimura2003magnetic}. It exhibits three magnetic transitions at 42~K, 28~K and 8~K. At 42~K,  Mn$^{3+}$ moments form  sinusoidal incommensurate AFM order\cite{quezel1977magnetic,kimura2003magnetic}. The magnetic order become commensurate and generates a spiral order in $\bf{bc}$ plane below 28~K, this accompanies ferroelectric order along the $\bf c$ axis which is switchable to $\bf a$ axis by applying magnetic field\cite{kimura2003magnetic}. One would expect that a 50:50 solid solution of TbFeO$_3$ and TbMnO$_3$ might be appropriate to combine the desirable properties of both compounds into one, especially considering the example of magneto-dielectric properties generated in YFe$_{1-x}$Mn$_x$O$_3$\cite{mandal_prl_107_2011spin}. Here, we present the spin-reorientation and magnetization reversal in TbMn$_{0.5 }$Fe$_{0.5 }$O$_3$ single crystals. Our studies indicate that strong anisotropic magnetic properties emerge in the half-doped crystals, disparate from both parent compounds. 
\section{Experimental}
Single crystals of TbMn$_{0.5}$Fe$_{0.5}$O$_3$ used in this study were grown by float-zone technique employing a four-mirror image furnace (FZ-T-10000-H-VI-VP from Crystal Systems Inc., Japan). Phase purity and crystal structure of the sample was confirmed by powder X-ray diffraction patterns from crushed crystal pieces. The powder diffractograms were obtained using a Phillips X'Pert High Score instrument using Cu$K\alpha$ radiation. Magnetic measurements were performed using a commercial Magnetic Property Measurement System from Quantum Design Inc. Specific heat was measured under zero applied field using a Physical Property Measurement System.
\section{Results and Discussion}
\begin{figure}[!t]
\centering
\includegraphics[scale=0.35]{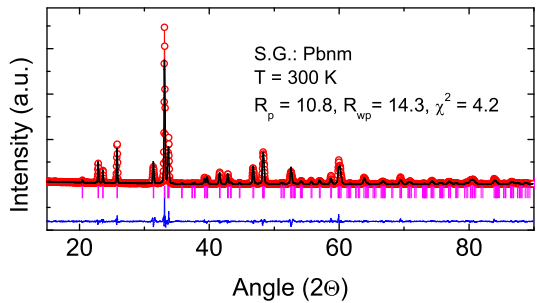}
\caption{(colour online) Powder x-ray diffraction pattern obtained on crushed single crystals of TbMn$_{0.5}$Fe$_{0.5}$O$_3$ at 300~K. Rietveld refinement results performed using $Pbnm$ space group are also presented. The goodness-of-fit was $\chi^2$ = 4.2.}
\label{xrd}
\end{figure}

\begin{figure}[!b]
\centering
\includegraphics[scale=0.40]{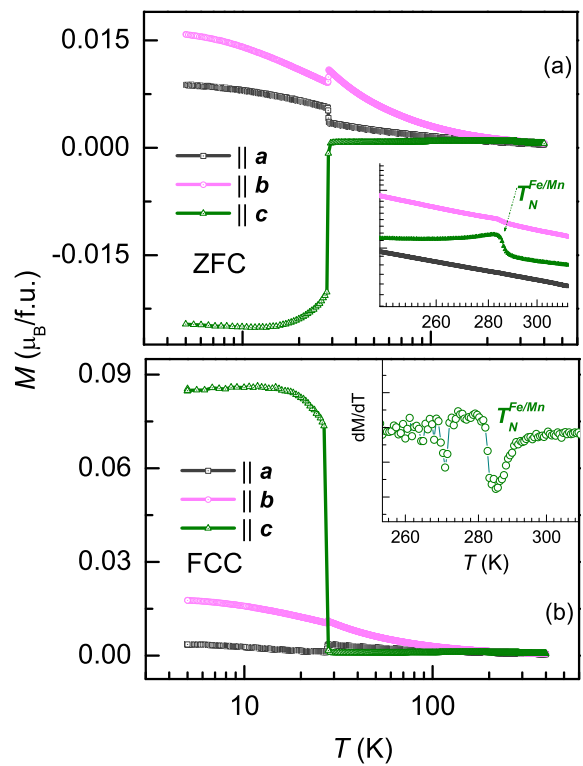}
\caption{(colour online) (a) ZFC curves at 100~Oe along three crystallographic axes, $\bf a$, $\bf b$ and $\bf c$. The inset shows a magnified view of the high temperature region where the transition at $T^\mathrm{Fe/Mn}_N $ = 286~K along $c$ axis is evident. (b) FCC curves along different axes. The inset magnifies the high temperature region around $T^\mathrm{Fe/Mn}_N $.}
\label{DCmag}
\end{figure}

\indent
The powder diffraction data obtained on crushed single crystal samples was refined using Rietveld method\cite{rietveld} implemented in FULLPROF code\cite{carvajal}. The refined structural parameters using $Pbnm$ space group are collected in Table~\ref{Table 1} and the refined pattern showing the observed, calculated data and the Bragg reflections is shown in Fig.~\ref{xrd}. The refinement was carried out in the space group $Pbnm$. The quality measures obtained are: $R_p$ = 10.8, $R_{wp}$ = 14.3, and $\chi^2$ = 4.2. A single crystal was oriented in three different crystallographic directions $\bf a$, $\bf b$ and $\bf c$ with the help of Laue camera for magnetic measurements.
\\
%
% % % % % % % % % % % % % % % % % % % % % % % % % % 
% % % % % % % % % % % % % % % % % % % % % % % % % %
%
%
\begin{table}[!b]
\begin{center}
\caption{\label{Table 1} The structural details of TbFe$_{0.5}$Mn$_{0.5}$O$_3$ obtained by refining powder x-ray diffraction data at 300~K using $Pbnm$ space group. The refined lattice parameters are $a$ = 5.311(7)\AA, $b$ = 5.697(2)\AA ~and $c$ = 7.539(3)\AA. }
\begin{tabular}{lllll}\hline\hline
Atom & Wyckoff pos.            &  x/a	                &  y/b	             & z/c 	\\ \hline
Tb      & $4c$			& -0.0155(2)		& 0.0729(3)		& 0.25		\\
Mn/Fe & $4b$      	& 0.5		            & 0		              & 0		\\
O1        & $4c$			& 0.1138(3)		    & 0.4629(2)		  &	0.25	\\
O2       & $8d$			& 0.6997(4)		& 0.3096(2)		  &	0.0586(1)	\\	\hline\hline
\end{tabular}
\end{center}
\end{table}	
\indent
Temperature dependent magnetization measurements were carried out on oriented single crystals along $\bf a$, $\bf b$ and $\bf c$ axes in a Quantum Design Inc. SQUID magnetometer. Fig~\ref{DCmag}(a) and (b) present the data in zero field cooled (ZFC) and field-cooled cooling (FCC) protocols at 100~Oe, respectively.  In TbMn$_{0.5}$Fe$_{0.5}$O$_3$, the first magnetic transition while cooling from 300~K is observed  near 286~K; denoted as $T^\mathrm{Fe/Mn}_N $ (highlighted in the inset of Fig~\ref{DCmag} (a)). Upon Mn-doping, the general trend observed in rare earth orthoferrites is a reduction of  $T_N$ and an enhancement of $T^\mathrm{Fe/Mn}_{SR}$\cite{nagata2001magnetic,chiang2011effect,hong2011continuously}. When the Mn doping level is near 50$\%$, $T_N$ and $T^\mathrm{Fe/Mn}_{SR}$ would be very close to each other\cite{nagata2001magnetic}. For example, in DyFe$_{0.5}$Mn$_{0.5}$O$_3$, the transition temperatures are $T^\mathrm{Fe/Mn}_N$ = 320~K and $T^\mathrm{Fe/Mn}_{SR}$ = 310~K respectively for $x$ = 0.5\cite{chiang2011effect}. The transition at 286~K in TbFe$_{0.5}$Mn$_{0.5}$O$_3$ is identified as a paramagnetic-to-AFM transition. The transition is more clearly shown in the inset of (b) where a plot of $dM/dT$ versus $T$ is plotted.
\\
%
%
% % % % % % % % % % % % % % % % % % % % % % % % % % 
% % % % % % % % % % % % % % % % % % % % % % % % % %
%
%
\indent
A second magnetic phase transition is observed for TbMn$_{0.5}$Fe$_{0.5}$O$_3$ at 28~K as a sharp feature along all the crystallographic directions (Fig~\ref{DCmag} (a) and (b)). However, this transition is most pronounced for the $\bf c$-axis. At this temperature denoted by $T^\mathrm{Fe/Mn}_{SR}$ to signify a spin-reorientation transition, the magnetic structure of TbMn$_{0.5}$Fe$_{0.5}$O$_3$ changes from $A_xG_yC_z$ (or simply $G_y$ because $A_x$ and $C_z$ components are very small)  to $G_xA_yF_z$. $G_y$ is chosen from the magnetic symmetry consideration \cite{yamaguchi1973magnetic} and the fact that there is no hysteresis loop at 250 K, below T$_N^{Fe/Mn}$ in any direction. In most $R$FeO$_3$ compounds, Fe$^{3+}$ orders in $G_xA_yF_z$ configuration below $T_N$\cite{PhysRevLett.1.3}. In TbMnO$_3$, Mn$^{3+}$ moments adopt an incommensurate spiral order at low temperature however, Tb also orders at very low temperatures below $\approx$ 7~K. It is natural to assume that the spin-reorientation transition in TbMn$_{0.5}$Fe$_{0.5}$O$_3$ would depend on the competition between the magnetic anisotropy energies of Mn$^{3+}$ and Fe$^{3+}$. 
At 28~K the anisotropy energy of Fe$^{3+}$ will be overcome by that of Mn$^{3+}$ ions and consequently, a $G_xA_yF_z$ configuration with a WFM component along $\bf c$- axis results. It must be noted that in TbMnO$_3$, the Mn$^{3+}$ ions order AFM along $\bf b$-axis below 28~K. Further support for the arguments about the nature of phase transitions at $T^\mathrm{Fe/Mn}_{N}$  and $T^\mathrm{Fe/Mn}_{SR}$  would follow from the hysteresis measurements presented in subsequent sections.
\\
\indent
In Fig~\ref{DCmag} (a), negative magnetization is observed for the magnetization data measured along $\bf c$-axis. The observed negative magnetization can be explained in terms of the domain wall pinning model described in the next paragraph. Spin reversal and negative magnetization are observed in SmFeO$_{3}$\cite{cao_nature_4_2014temperature} or NdFeO$_3$\cite{yuan2013spin} where, a spontaneous spin reversal of the Fe/Mn and $R$ sub-lattices leads to the first-order jump in magnetization. The phenomena of negative magnetization is well documented in recent reviews\cite{kumar_physrep_2014phenomenon} and note that care must be exercised in explaining its origins. Rare earth ordering in $R$FeO$_3$ or $R$MnO$_3$ is reflected as a anomaly in magnetization below 4~K along $\bf a$ and $\bf b$ axis\cite{kimura_prb_71_2005magnetoelectric,bellaiche2012simple,PhysRevLett.1.3}. In TbMn$_{0.5}$Fe$_{0.5}$O$_3$ a weak, broad cusp is seen at 12~K in the magnetization measured along $\bf c$ axis. However, this cannot be accounted in terms of Tb ordering as Tb usually orders below 8.5~K with the spins confined in the $\bf{ab}$ plane\cite{quezel1977magnetic}. Curie-Weiss analysis of the magnetization data along three axes reveals FM interaction along $\bf b$-axis and AFM components along the other two axes as listed in Table~\ref{Table 2}. 
\\
%
%
% % % % % % % % % % % % % % % % % % % % % %
\begin{figure}[!t]
\centering
\includegraphics[scale=0.40]{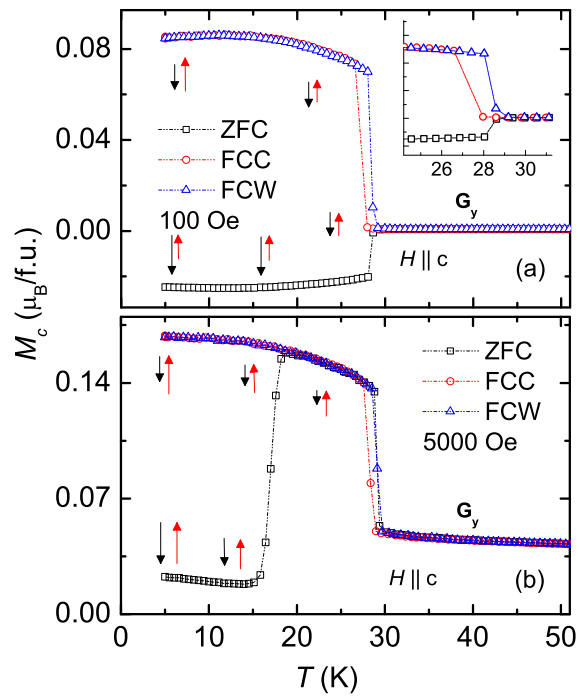}
\caption{(colour online) (a) ZFC, FCC and FCW  measurements at 100~Oe along $\bf c$ axis. (b) represents same set of measurements performed at 5000~Oe. The data shown is limited upto 50~K. Inset shows the thermal hysteresis seen at $T^\mathrm{Fe/Mn}_{SR}$ transition. Data collected upon warming are shown in blue, and upon cooling in red. A schematic representation of the magnetic domain strength along and  opposite to the applied field are drawn as red and black arrows, respectively.}
\label{DCmag2}
\end{figure}
% % % % % % % % % % % % % % % % % % % % % % 
%
%
 \indent 
FCC and field-cooled warming (FCW) magnetization measurements performed at 100~Oe along  $\bf c$-axis (denoted as $M_c$) after oscillating the field from  30~kOe and making the field minimum and positive are presented in Fig~\ref{DCmag2} (a) along with the corresponding ZFC curve. $M_c$ exhibits contrasting temperature dependence in ZFC and FCC data. Clear irreversibility is seen between the FCC and FCW curves close to $T^\mathrm{Fe/Mn}_{SR}$ (inset of fig~\ref{DCmag2} (a)), where $M_c$ changes from  -0.02 to  +0.0017 $\mu_B$/f.u. The transition width ($\Delta T $) is about 2~K. The negative magnetization for $M_c$ could be explained in terms of domain wall pinning. Since Tb moment have no FM component along $\bf c$-axis for either of the parent compounds, a competition between 4$f$ moment and 3$d$ moments could be excluded. During  ZFC measurements, magnetic moments are likely to align according to the anisotropy and internal magnetic exchanges and magnetic domains can form along different directions. Depending upon the initial configuration of the domains and strength of domain wall pinning even an overall negative magnetization could result. When the temperature increases in the ZFC measurements more and more domains align towards the field but still with an overall negative magnetization. This explains the gradual increase in  $M_c$ till $T^\mathrm{Fe/Mn}_{SR}$. At 28~K the spin structure reorient to $G_y$. In FCW measurements, most of the domains are aligned towards the applied field giving a positive total magnetic moment. With increase in temperature, $M_c$ shows  moderate decrease up to $T^\mathrm{Fe/Mn}_{SR}$ due to the decrease in domain strength of those domains oriented opposite to the field. At $T^\mathrm{Fe/Mn}_{SR}$, $M_c$ shows a sudden fall  from  0.07 to  0.0017 $\mu_B$/f.u. The sudden fall could be because of the spin-reorientation back to the $G_y$ configuration. Apart from the usual spin-reorientation, a total magnetization reversal is observed in NdFeO$_3$ and SmFeO$_3$\cite{yuan2013spin} similar to the present case. The competition between rare earth 4$f$ sub-lattice and transition element 3$d$ sub-lattice in $R$FeO$_3$ plays a major role in the spin-reorientation transition\cite{yuan2013spin}.
 \\
 %
 %
%% % % % % % % % % % % % % % % % % % % % % % % %
\begin{table}[!t]
\caption{\label{Table 2} Effective paramagnetic moments and Curie-Weiss temperatures of TbMn$_{0.5}$Fe$_{0.5}$O$_3$ extracted from performing Curie-Weiss analysis of the magnetization data along the three different axes.}
\begin{tabular}{llll}\hline\hline
                                                                              & $\bf a$            & $\bf b$          & $\bf c$ \\ \hline\hline
$\mu_\mathrm{eff}$ ($\mu_\mathrm{B}$/f.u.)   & 9.35(3)              & 10.75(1)           &  13.11(3)      \\        
$\Theta_{CW}$ (K)                                        & -5.5(2)              & 15.3(4)            & -227.5(2)       \\ \hline\hline 
\end{tabular}
\end{table}	
% % % % % % % % % % % % % % % % % % % % % % % 
%
%
% % % % % % % % % % % % % % % % % % % % % % % % % %
\begin{figure}[!h]
\centering
\includegraphics[scale=0.43]{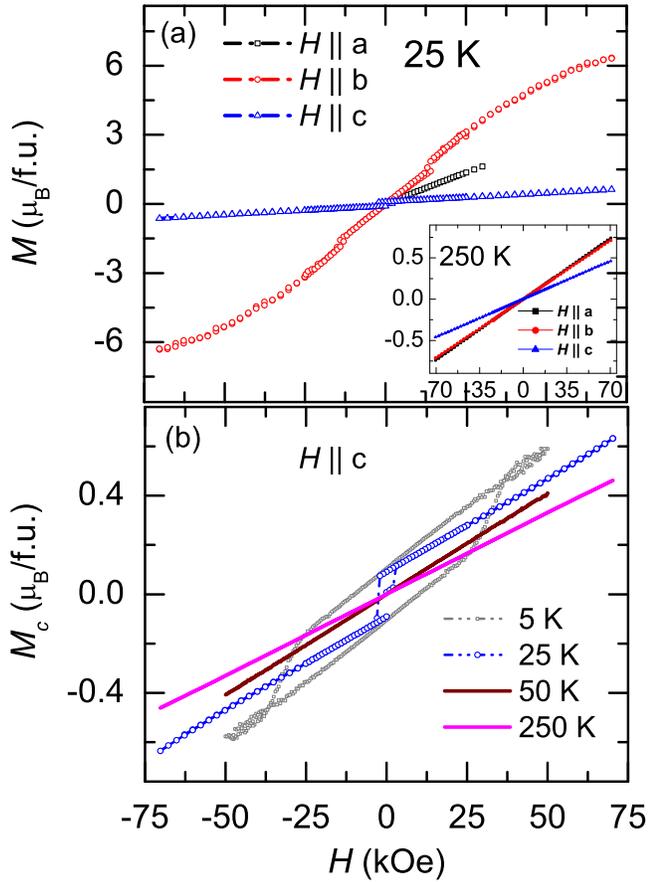}
\caption{(colour online) (a) Field-scans of magnetization, $M (H)$, along $a$, $b$ and $c$ axis at 25~K and 250~K (inset). Anomalies (step-like) are observed in the $\bf b$-axis data at $\approx$ 25~kOe. (b) $M (H)$ along $\bf c$ axis at different temperatures. A weak component of ferromagnetism is revealed at low temperatures.}
\label{MH}
\end{figure}
% % % % % % % % % % % % % % % % % % % % % % % % % %
%
%
\indent 
$M_c$ was measured at higher applied magnetic field to check if the spin-reorientation transition changes with the magnetic field and the results are presented in Fig~\ref{DCmag2} (b). At higher applied field, negative $M_c$ is not observed, but the overall nature of $M_c$ is similar to that measured at 100~Oe. In case of ZFC, even though domains with a resultant orientation opposite to the applied filed are present, an overall positive moment is observed. This is due to the fact that at a higher applied field, more domains will orient towards the field.  With increase in temperature more and more domains align towards the field and at a critical temperature ($\sim$16 K) most of the domains are orient towards the field. In the FCC and FCW measurements most of the domains are already oriented along the field direction and with increase in temperature the strength of the domains decreases. This explains the gradual decrease of moment till $T^\mathrm{Fe/Mn}_{SR}$.  The spins reorient from $G_xA_yF_z$ configuration to $G_y$ configuration near 28~K, which explains the sudden fall in $M_c$. Interestingly, in the parent TbMnO$_3$, AFM incommensurate-commensurate transition responsible for the ferroelectric phase is reported to happen exactly at 28~K \cite{kimura2003magnetic}. The sinusoidal AFM ordering is along $\bf b$ axis in TbMnO$_3$. In $R$Fe$_{0.5}$Mn$_{0.5}$O$_3$, the anisotropy energy of Fe moments is overcome by that of Mn moments at a specific temperature\cite{nagata2001magnetic,chiang2011effect}. Due to high magnetic anisotropy of Mn moments, the Fe moments will align towards the direction of Mn moments. This normally changes the AFM direction from $\bf a$ axis to $\bf b$ axis at $T^\mathrm{Fe/Mn}_{SR}$\cite{nagata2001magnetic}. 

\indent 
Field-scans of magnetization along the $H \| \bf a$ , $H \| \bf b$ and $H \| \bf c$ are shown in fig~\ref{MH} (a) at 25~K while the inset shows that at 250~K. No hysteresis is observed along  $H \| \bf a$, $H \| \bf b$ or $H \| \bf c$ at 250~K. This fact confirms that the transition at 286~K is an AFM ($A_xG_yC_z$ or simply $G_y$) transition. At 25~K, just below the $T^\mathrm{Fe/Mn}_{SR}$ transition, a hysteresis loop opens only for $M_c$ (this is clear from the panel (b)). This indicate that at $T^\mathrm{Fe/Mn}_{SR}$ magnetic structure changes from $G_y$ to $G_xA_yF_z$  ordering with WFM along $\bf c$-axis.  It is understood that weak ferromagnetism is not present along $\bf a$ or $\bf b$ direction as suggested for other orthoferrites. Magnetic moment of $\sim$ 7.5 $\mu_B$/f.u. is observed at high field for magnetization measured along $b$-axis. The high value of moment along $\bf a$  and $\bf b$-axes is from a possible $F_xC_y$-like ordering of Tb$^{3+}$ moments as seen in the parent compound, TbFeO$_3$\cite{tejada1995quantum}. A first-order spin-flip transition occurs at $H_c \pm$ 26~kOe evident in the $\bf b$-axis data. In the magnetization along $\bf a$, no hysteresis is detected even at 2~K (not shown). Note that Table-\ref{Table 2} revealed predominant AFM interactions present along $\bf a$ axis. 
Detailed isothermal measurements are performed along $\bf c$ and are exhibited in fig~\ref{MH} (b). Although the magnetization along $\bf a$ or $\bf b$ show no hysteresis, the magnitude of $M_a \approx$ 5$M_c$ while $M_b \approx$ 10$M_c$. In  $H \| \bf c$ measurements, a clear hysteresis loop is observed at 5~K. Even at 50~kOe, $M_c$ is not saturated which indicates the underlying antiferromagnetic state of the compound\cite{bazaliy2004spin}.       
\\
%
%
 % % % % % % % % % % % % % % % % % % % % % %
 \begin{figure}[!h]
 \centering
 \includegraphics[scale=0.50]{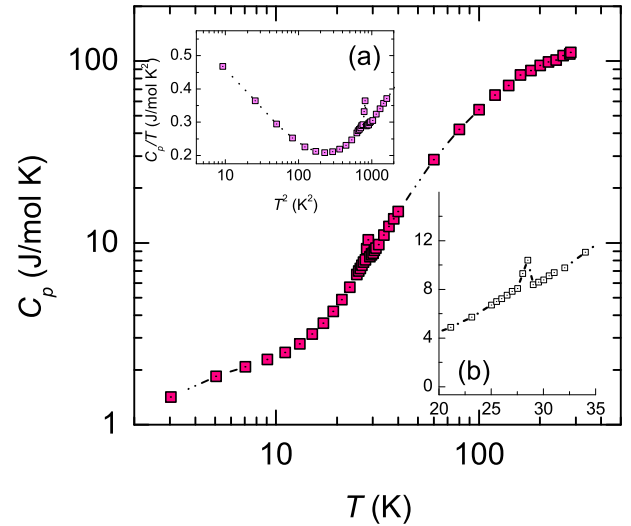}
 \caption{(colour online) Specific heat of TbFe$_{0.5}$Mn$_{0.5} $O$_3$ single crystal at zero applied magnetic field. The inset (a) is a plot of $C_p/T$ versus $T^2$. The inset (b) shows a magnified view near $T^\mathrm{Fe/Mn}_{SR}$.}
 \label{Cp}
 \end{figure}
 % % % % % % % % % % % % % % % % % % % % % % 
 %
 %
 Finally, the specific heat of TbFe$_{0.5}$Mn$_{0.5} $O$_3$ single crystal measured using heat-pulse method in the PPMS is presented in the main panel of Fig~\ref{Cp} in log-log scale. Spin-reorientation transition at $T^\mathrm{Fe/Mn}_{SR}$ is clearly evident in the plot as a sharp cusp. The broad magnetic $G_x$ transitions at $T^\mathrm{Fe/Mn}_{N}$ is not visible as sharp transition but rather as a broad feature in the derivative, $dC_p/dT$ (not shown). In ErFeO$_3$, a distinct enhancement in specific heat value in the spin reorientation regime is reported \cite{chaudhury2009lattice}. Inset of Fig~\ref{Cp} (a) shows the plot of $C_p/T$ over $T^2$. Below the transition at $T^\mathrm{Fe/Mn}_{SR}$ which shows as a peak, a gradual decrease in $C_p/T$ is observed and then eventually a sharp increase. The sharp increase at low temperature is due to the rare-earth ordering. In TbFeO$_3$, Tb is reported to order at 3.2 K \cite{de1968chaleurs}. The linear region below the transition was used to fit the expression $C_p$ = $\beta_3T^{3}$ + $\beta_5T^{5}$. From this fit, the value for $\beta$ is obtained to be used in the relation $\Theta_D$ = ($\frac{12p\pi^4R}{5\beta_3}$)$^{1/3}$; where $p$ is the number of atoms in the unit cell and $R$ is the universal gas constant. $\Theta_D$ is estimated to be $\approx$ 480~K, matching with the value reported for TbFeO$_3$ by Parida et al. \cite{parida2008heat}. A magnified view of the temperature region near $T^\mathrm{Fe}_{SF}$ is presented in the inset (b).
\section{Conclusions}
\indent 
The magnetic properties of single crystal TbFe$_{0.5}$Mn$_{0.5} $O$_3$ along different crystallographic axes are investigated in detail. Prominent anisotropy effects are observed. At $T^\mathrm{Fe/Mn}_{N}$ = 286~K antiferromagnetic $A_xG_yF_z$ (or simply $G_y$) ordering is observed which reorient to  $G_xA_yF_z$ configuration below $T^\mathrm{Fe/Mn}_{SR}$ = 28~K. The magnetic properties of the mixed-solution TbFe$_{0.5}$Mn$_{0.5} $O$_3$ are found to be entirely different from both the end compounds, TbMnO$_3$ and TbFeO$_3$. Strong competition between the magnetic anisotropy energy of Fe and Mn moments make this compound magnetically complicated. A component of weak ferromagnetism is clearly revealed along $\bf c$ axis which also presents negative magnetization in low fields. Detailed neutron diffraction measurements and subsequent inelastic scattering experiments using single crystals are planned in order to shed more light in to the complex $H-T$ phase diagram of TbFe$_{0.5}$Mn$_{0.5} $O$_3$.
\\ \\
Authors HN and SE wish to acknowledge the Department of Science and Technology (DST) for financial support. HSN acknowledges FRC/URC for a postdoctoral fellowship. AMS thanks the SA NRF (93549) and UJ URC/FRC for financial assistance.
% 

% REFERENCES
%\bibliographystyle{aipnum4-1}
%\bibliography{TFMO}
%
%
%
\end{document}